\documentstyle[emulateapj,epsf]{article}

\def\agt{\mathrel{\raise.3ex\hbox{$>$}\mkern-14mu\lower0.6ex\hbox{$\sim$}}}
\def\alt{\mathrel{\raise.3ex\hbox{$<$}\mkern-14mu\lower0.6ex\hbox{$\sim$}}}

\newcommand{\beq}{\begin{equation}}
\newcommand{\eeq}{\end{equation}}
\newcommand{\beqn}{\begin{eqnarray}}
\newcommand{\eeqn}{\end{eqnarray}}
\newcommand{\pa}{\partial}

\begin{document}

\title{Stability of rigidly rotating relativistic stars
with soft equations of state against gravitational collapse} 

\author{Masaru Shibata \altaffilmark{1}}

\affil
{\altaffilmark{1} 
Graduate School of Arts and Sciences, 
University of Tokyo,
\break
Komaba, Meguro, Tokyo 153-8902, Japan}



\begin{abstract}

We study secular stability against a quasi-radial oscillation for 
rigidly rotating stars with soft equations of state in general relativity. 
The polytropic equations of state with polytropic index $n$ 
between 3 and 3.05 are adopted for modeling the rotating stars.
The stability is determined in terms of the turning-point method. 
It is found that (i) for $n \agt 3.04$, 
all the rigidly rotating stars are unstable against 
the quasi-radial oscillation and (ii) for $n \agt 3.01$, the 
nondimensional angular momentum parameter $q \equiv cJ/GM^2$ (where 
$J$, $M$, $G$, and $c$ denote the angular momentum, the gravitational mass, 
the gravitational constant, and the speed of light, respectively) 
for all marginally stable rotating stars is larger than unity.
A semi-analytic calculation is also performed, and good agreement
with the numerical results is confirmed. The final outcome after
axisymmetric gravitational collapse of rigidly 
rotating and marginally stable massive stars with $q > 1$ is predicted,
assuming that the
rest-mass distribution as a function of the specific angular momentum
is preserved and that the pressure never halt the collapse.
It is found that even for $1 < q \alt 2.5$, a black hole may be 
formed as a result of the collapse,
but for $q \agt 2.5$, the significant angular momentum will
prevent the direct formation of a black hole. 

\end{abstract}




\section{Introduction}

At the final stage of their evolution, 
massive stars with initial mass $\agt 8M_{\odot}$ form an iron core.
Because of high temperature and density, 
the photodissociation of the iron to lighter elements
proceeds rapidly (Shapiro \& Teukolsky 1983).
In the photodissociation, the thermal pressure is consumed, decreasing 
the adiabatic index $\Gamma$ below $4/3$.
Consequently, the core is destabilized, 
resulting in collapse to a neutron star or a black hole. 

Recent numerical simulations of the collapse of 
primordial molecular clouds have suggested that the first generation 
of stars contains many massive members of mass 
$\sim 100$--$1000M_{\odot}$ (Bromm et al. 1999; 
Abel et al. 2000; Nakamura \& Umemura 2001). 
Such massive stars form large helium cores that reach carbon ignition. 
It is known that after helium burning, cores of such stars 
encounter the electron-positron pair creation instability, 
bringing $\Gamma$ below $\sim 4/3$ and triggering the collapse 
(Barkat et al. 1967; Bond et al. 1984; Kippenhahn \& Weighrt 1990; 
Fryer et al. 2001).  For sufficiently massive stars, of mass 
$M \agt 260 M_{\odot}$, such unstable cores form a black hole, 
while for less massive cases, the outcome is a pair-unstable 
supernova (Fryer et al. 2001). 

These examples illustrate that
the gravitational collapses set in when the
stability of progenitors changes, and hence, 
a star marginally stable against quasi-radial oscillations may be
regarded as an initial condition of the collapse.
Massive stars in nature are rapidly rotating in general (Bond et al. 1984).
Thus, determining the quasi-radial stability for rotating stars 
is an important subject in astrophysics for elucidating 
a plausible initial condition for the stellar collapse. 

In many cases, one simply assumes that progenitors of neutron stars or
black holes start collapsing
when their adiabatic index $\Gamma$ decreases below a well-known
critical value, 4/3, against a quasi-radial instability. The number $4/3$ 
is the correct value only when the progenitor is not rotating.
If the progenitor is rotating, it is not the valid criterion, 
since angular momentum has the effect of stabilizing the star 
against gravitational collapse. In particular, the number 4/3
would not be a good indicator of the instability for 
pair-unstable collapse, since the progenitor stars are likely
to be rapidly rotating (Bond et al. 1984, Fryer et al. 2001). 

According to a Newtonian theory, 
slowly rotating stars that satisfy the following relation are 
unstable against a quasi-radial oscillation (Tassoul 1978; see also \S 2): 
\beq
3\Gamma - 4 - 2 (3\Gamma-5)\beta < 0. \label{eq1}
\eeq
Here, 
\beq
\beta \equiv {T \over W},
\eeq
and $T$ and $W$ are rotational kinetic and gravitational 
potential energies, respectively. In the following, we restrict our attention 
to the rigidly rotating case for simplicity. 
Based on our numerical analysis in Newtonian gravity,
the maximum value of $\beta$ for rigidly rotating stars
with polytropic equations of state, which is achieved when
the velocity at the equatorial surface is equal to the
Kepler velocity (i.e., at the mass-shedding limits), 
is approximately written as 
\beq
\beta_{\rm max} \approx 0.00902 + 0.124 \biggl( \Gamma-{4\over 3}\biggr).
\label{eq2}
\eeq
Combining equations (\ref{eq1}) and (\ref{eq2}), we find 
that for $\Gamma \alt 1.328$, all the rigidly rotating 
stars are unstable against a quasi-radial oscillation
in Newtonian gravity.  However, even for $1.328 \alt \Gamma < 4/3$,
the stars can be stabilized by the effect of rotation. 

General relativistic gravity destabilizes the rapidly rotating
stars even in the case of $\Gamma > 1.328$. One of the
purposes of this paper is to numerically determine the criteria
for the onset of the instability. 
In particular, we focus on the secular stability of rigidly
rotating stars with soft equations of state in equilibrium 
against a quasi-radial oscillation. The stability is determined
using the turning-point method (Friedman et al. 1988). 
To model the rotating stars, we adopt the polytropic equations
of state with $n \agt 3~(\Gamma \alt 4/3)$.

{}From a general relativistic point of view, 
it is interesting to ask how large the nondimensional
angular momentum parameter $q \equiv J/M^2$ can be for the marginally stable
stars that may be plausible progenitors of gravitational collapse. 
Axisymmetric hydrodynamic simulations in general relativity 
(Nakamura 1981; Stark \& Piran 1985; Piran \& Stark 1986; 
Nakamura et al. 1987; Shibata 2000)
have provided many numerical results that show that for $q > 1$,
a black hole is not formed after the collapse. 
That is, the value of $q$ may be a good indicator in predicting the 
final fate of the stellar collapse of massive objects. 
As Baumgarte and Shapiro (1999) indicate, 
the value of $q$ for a marginally stable star with $\Gamma=4/3$ 
is very close to unity. Cook et al. (1994) show that 
the critical value of $q$ increases 
with decreasing of $\Gamma$ for $4/3 < \Gamma \alt 5/3$. 
This implies that for $\Gamma < 4/3$, the critical value is likely
to be larger than unity, even for the case of $4/3-\Gamma \ll 1$.
The second purpose of this paper is to confirm this fact.

The third purpose is to predict the outcomes of gravitational
collapse of the marginally stable stars. 
As mentioned above and confirmed below,
the value of $q$ for the marginally stable stars
with $\Gamma < 4/3$ can be larger than unity. 
This suggests that a rotating star with $\Gamma < 4/3$ of
mass large enough collapse to a black hole in the nonrotating case 
may not form a black hole because of its significant effect of
the angular momentum. 
In the analysis, we assume that 
(i) the collapse proceeds in an axisymmetric manner,
(ii) the rest-mass distribution as a function of
the specific angular momentum is preserved, and
(iii) the pressure never halt the collapse.
We indicate that the criterion for no black hole formation,  
$q>1$, which has been suggested so far, 
is not very good for rigidly rotating stars with 
the soft equations of state. We predict 
that even for $1 < q \alt 2.5$, a black hole may be formed
as a result of gravitational collapse of the rigidly 
rotating and marginally stable stars. However, for $q \agt 2.5$, 
the significant angular momentum will prevent the collapsing stars
from directly forming a black hole.

The paper is organized as follows. 
In \S 2, we present a semi-analytic calculation for determining 
the stability of rotating stars against a quasi-radial oscillation. 
In \S 3, the secular stability of rotating stars in general relativity 
is numerically determined. 
In \S 4, we predict the final outcomes of the
stellar collapse for very massive stars, 
assuming that the initial condition is a marginally stable
star, as determined in \S 3. 
Section 5 is devoted to a summary.  Throughout this paper,
the pressure is given by the polytropic equation of state as 
\beq
P= K \rho^{\Gamma},~~\Gamma=1+{1 \over n}, 
\eeq
where $P$ is the pressure, $\rho$ the baryon rest-mass density,
$K$ the polytropic constant, and $n$ the polytropic index, which
is chosen in the range between 3 and 3.05.
We adopt the geometrical units $G=c=1$, where $G$ and $c$ 
denote the gravitational constant and the speed of light, 
respectively.

\section{Semi-analytic exploration for stability}

To semi-analytically determine the marginally stable rotating
stars with the polytropic equations of state with 
$n \agt 3$, we follow the method
described by Zel'dovich \& Novikov (1971) and 
Shapiro \& Teukolsky (1983) and write the total energy as
the sum of the internal energy, the Newtonian potential energy,
the rotational kinetic energy (in Newtonian order),
and a post-Newtonian correction. These terms can be written
\beqn
E&=&k_1 K M \rho_c^{1/n}-k_2 M^{5/3}\rho_c^{1/3} \nonumber \\
&& +k_5 J^2 M^{-5/3}\rho_c^{2/3}-k_4 M^{7/3} \rho_c^{2/3},\label{eq21}
\eeqn
where $k_1$, $k_2$, $k_4$, and $k_5$ are structure
constants that are constructed from 
Lane-Emden functions (Shapiro \& Teukolsky 1983) and listed in Table 1. 
$M$ is the total mass and $\rho_c$ the central density. 

To derive equation (\ref{eq21}), we 
assume that the stars are spherical and rigidly rotating.
In the treatment by Zel'dovich \& Novikov (1971) and Shapiro \& Teukolsky
(1983), the authors take into account a nonsphericity, assuming that
a spherical surface of constant density in the nonrotating case
transforms into a spheroidal surface enclosing the same volume.
This approximate method is likely to be good for stars with 
very stiff equations of state with $n \sim 0$.
However, this is not the case for the stars with $n \agt 3$, 
since the density profile is not as uniform as that for $n=0$.
Actually, in their approximate treatment, the value of $\beta$
for given values of $\rho_c$ and $M$ is 
significantly overestimated for $n \sim 3$. In the case 
of the soft equations of state, the shape of the central region, 
around which most of the mass is concentrated, 
is nearly spherical even at mass-shedding limits. 
Therefore, we neglect the nonsphericity here. 
In this treatment, the value of $J$ is considered an input quantity.
Note also that the second post-Newtonian term is omitted in the
present treatment, in contrast to those by Zel'dovich \& Novikov (1971)
and Baumgarte \& Shapiro (1999). This is because the compactness 
of stable stars with a soft equation of state is very small, and 
hence, the second post-Newtonian correction produces 
a very small effect on the equilibrium and stability, 
as illustrated by Baumgarte \& Shapiro (1999) for $n=3$. 

Taking the first derivative of equation (\ref{eq21}) with respect to
the central density yields a condition for equilibrium, 
\beqn
0={\pa E \over \pa x}
&=&{3 \over n} k_1 K M x^{(3-n)/n}-k_2 M^{5/3} \nonumber \\
&& +2 k_5 J^2 M^{-5/3}x - 2 k_4 M^{7/3} x,\label{eq22}
\eeqn
where $x\equiv \rho_c^{1/3}$. For a stable equilibrium,
the second derivative of equation (\ref{eq21}) has to be positive.
Therefore, a root of the second derivative marks the onset of
a quasi-radial instability: 
\beqn
0={\pa^2 E \over \pa x^2}
&=&{3 \over n}\biggl({3 \over n}-1\biggr)
k_1 K M x^{(3-2n)/n} \nonumber \\
&& +2 k_5 J^2 M^{-5/3} - 2 k_4 M^{7/3}.\label{eq23}
\eeqn
By combining equations (\ref{eq22}) and (\ref{eq23}), we derive 
a relation between $M$ and $x$ for the marginally stable stars as 
\beqn
M=\biggl[{3(2n-3)k_1 \over n^2k_2}K x^{(3-n)/n}\biggr]^{3/2}. 
\label{eq24}
\eeqn
For $n=3$, $MK^{-3/2}$ is a constant as 4.5548 irrespective of $x$ 
(Baumgarte \& Shapiro 1999). 
For $n > 3$, $M$ decreases with increasing $\rho_c$ for
a fixed value of $K$, and 
in the limit $\rho_c \rightarrow 0$, $M \rightarrow \infty$. 

From equations (\ref{eq23}) and (\ref{eq24}),
$J/M^2$ for the marginally stable stars (defined as $q_{\rm mar}$)
can be written as a function of the density: 
\beqn
q_{\rm mar}^2={k_4 \over k_5} + {n^2(n-3)k_2^2 \over 6(2n-3)^2k_1 k_5}
(\rho_c K^n)^{-1/n}.\label{eqqq}
\eeqn
For $n=3$, $q_{\rm mar}$ is a constant as 0.873 and for $n > 3$,
$q_{\rm mar}$ decreases with increasing $\rho_c$.
In the limit $\rho_c \rightarrow 0$, $q_{\rm mar} \rightarrow \infty$. 
As we show in \S 3, the maximum value of 
$\rho_c$ along the sequence of the marginally stable stars
is reached at mass-shedding limits. At that point, $q_{\rm mar}$ reaches its 
minimum ($q_{\rm min}$). Thus, the value of $q_{\rm mar}$
for the marginally stable stars is always larger than $q_{\rm min}$
for $n > 3$. 

Using equation (\ref{eq22}), 
equation (\ref{eq23}) can be rewritten as 
\beqn
0=3\Gamma-4 - 2(3\Gamma-5)\biggl({k_5 \over k_2}q^2 -
{k_4 \over k_2}\biggr)M^{2/3}\rho_c^{1/3}. \label{eq25}
\eeqn
In the present approach, 
the second term is related to $\beta(=T/W)$ by 
\beq
\beta= {k_5 \over k_2} q^2 M^{2/3}\rho_c^{1/3}. 
\eeq
Using this relation, equation (\ref{eq25}) is rewritten as  
\beqn
0=3\Gamma-4 - 2(3\Gamma-5)\biggl(\beta -
{k_4 \over k_2}M^{2/3}\rho_c^{1/3}\biggr). \label{eq26}
\eeqn
The term $M^{2/3}\rho_c^{1/3}$ is related to the compactness parameter
$M/R_s$, where $R_s$ is the radius of the spherical star, as follows:
\beq
{M \over R_s} = \alpha M^{2/3}\rho_c^{1/3}.\label{eq27}
\eeq
Here, the values of $\alpha$ are determined by the Lane-Emden
function and listed in Table 1. 
Using this relation, equation (\ref{eq26}) is written as 
\beqn
0=3\Gamma-4 - 2(3\Gamma-5)\beta -\alpha' {M \over R_s},
\label{eq28}
\eeqn
where $\alpha'=2(5-3\Gamma)k_4/(\alpha k_2)$ which is in the
range between 6.747 and 7.137 for $\Gamma=4/3$--1.328. 
Equation (\ref{eq28}) is a familiar formula often presented in 
the standard textbooks such as Tassoul (1978).
For the case in which $\beta=0$, the equation agrees with the formula
derived by Chandrasekhar (1964) for $n=3$. 

\section{Numerical analysis for stability} 

In \S 2, we determined sequences of the marginally stable stars for
$n \agt 3~(\Gamma \alt 4/3)$ in terms of a semi-analytic calculation.
As shown in this section, they indeed give good
approximate sequences for the marginally stable rotating stars.
However, they are not exact sequences after all. 
To determine them precisely, numerical computations are necessary. 

Another drawback in the semi-analytic calculation
is that the maximum value of the angular velocity for a
given value of $J$ is not determined precisely.
For rigidly rotating stars,
the velocity should be smaller than the Kepler velocity
at the equatorial surface. This restricts the allowed region
for the rigidly rotating stars: A sequence 
of stars at mass-shedding limits divides 
the regions where the rigidly rotating stars exist and where they do not. 
To determine the sequence at the mass-shedding limits, 
the configuration of the stars has to be accurately computed. 
Thus, numerical computation is inevitable. 

\subsection{Basic equations}

To study the secular stability of rotating stars against
a quasi-radial oscillation, the equilibrium solutions
in general relativity are computed. We write the energy momentum 
tensor of the Einstein equation for an ideal fluid as 
\beqn
T^{\mu\nu}=\rho h u^{\mu} u^{\nu} + P g^{\mu\nu},
\eeqn
where $u^{\mu}$ is the four velocity, 
$h \equiv 1+ \varepsilon + P/\rho$ 
is the enthalpy, $\varepsilon$ is the specific internal energy, and 
$g^{\mu\nu}$ is the spacetime metric.
As in \S 2, the polytropic equations of state are adopted. 
Using the first law of the thermodynamics,
the specific internal energy $\varepsilon$ in the polytropic
equations of state is written as
\beq
\varepsilon={n P \over \rho}. 
\eeq
As in \S 2, we pay attention only to rigidly rotating stars 
setting the angular velocity $\Omega \equiv u^{\varphi}/u^t$ as a 
constant.

With the polytropic equation of state, 
physical units enter the problem only through the polytropic constant 
$K$, which can be chosen arbitrarily or else completely scaled out of 
the problem.  Indeed, $K^{n/2}$ has units of length, time and mass, 
and $K^{-n}$ has units of density in the geometrical units. Thus, 
in the following, we only show non-dimensional quantities, 
which are rescaled by $K$. In other words, we adopt the
units with $c=G=K=1$. 

Following Butterworth \& Ipser (1976), the line element is written as
\beqn
ds^2 &=& -e^{2\nu} dt^2
+ B^2 e^{-2\nu} r^2 \sin^2\theta (d\varphi - \omega dt)^2 \nonumber \\
&& +e^{2\zeta-2\nu}(dr^2 + r^2 d\theta^2), 
\eeqn
where $\nu$, $B$, $\omega$, and $\zeta$ are field functions. The 
first three obey elliptic-type equations in axial symmetry, and 
the last one an ordinary differential equation. These equations 
are solved using the same method as that 
described by Shibata \& Sasaki (1998). 

The total baryon rest-mass $M_*$, Komar mass (gravitational mass) $M$,
proper mass $M_{\rm p}$, Komar angular momentum $J$,
rotational kinetic energy $T$, and gravitational potential
energy $W$ are defined from the energy momentum tensor or matter
variables as
\beqn
M_* &=& 2\pi \int \rho u^t B e^{2\zeta-2\nu} r^2 dr d(\cos\theta), \\ 
M &=&2\pi \int (-2T_t^{~t}+T_{\mu}^{~\mu}) B e^{2\zeta-2\nu}
r^2 dr d(\cos\theta),~~\\ 
M_{\rm p} &=&2\pi \int \rho u^t (1+\varepsilon)
B e^{2\zeta-2\nu} r^2 dr d(\cos\theta), \\ 
J  &=&2\pi \int \rho h u^t u_{\varphi} B e^{2\zeta-2\nu}
r^2 dr d(\cos\theta),\\ 
T&=&{1 \over 2} J \Omega,\\
W&=& M_{\rm p} - M + T, 
\eeqn
where $W >0$. 
From these quantities, the well-known
nondimensional parameters are defined as
$\beta \equiv T/W$ and $q \equiv J/M^2$. 
The ADM mass $M_{\rm ADM}$ is defined from the asymptotic
behavior of $\nu$ as
\beq
M_{\rm ADM} = -\lim_{r\rightarrow \infty} \nu r.
\eeq
For the stationary spacetime, $M_{\rm ADM}$ is equal to $M$ (Beig 1978). 
Thus, accuracy of the numerical solutions can be measured
checking the relation $M=M_{\rm ADM}$. 

In addition to these quantities, we often refer to the central density, 
$\rho_c$, which is used to specify a rotating star for a given set
of $\beta$ and $M$, and to the equatorial circumferential radius, $R$, 
by which a compactness parameter is defined as $M/R$.

\subsection{Analysis for secular stability}

The secular stability for rigidly rotating stars against
quasi-radial oscillations can be determined by a turning-point 
method as established by 
Friedman et al. (1988) and subsequently used by Cook et al. (1992, 1994). 
According to the turning-point theorem, 
a change of the sign of $dM/d\rho_c$ along a curve of a
constant value of $J$ indicates the change of a secular stability. 
Thus, in the numerical computation, 
curves of constant values of $J$ are computed and
plotted in the plane composed of $M$ and $\rho_c$ for determining 
the turning points.

To determine the region for the stable stars, in the present case, 
one should compute $dM/d\rho_c$ along curves of a constant value of $J$. 
For $(dM/d\rho_c)_J >0~(<0)$, the stars are stable (unstable) 
against gravitational collapse. From this fact,
one can distinguish the stable region from the unstable one.

\subsection{Numerical results}

Numerical computation was carried out using the spherical polar
coordinates $(r, \theta, \varphi)$. For a solution of elliptic-type
equations in axial symmetry, a uniform grid for $r$ and $\cos\theta$
is adopted with the typical grid size $(N_r, N_{\theta})=(1000, 160)$. 
To investigate convergence for the location of the turning points, the
computations were also performed with $N_r=500$, 750, and 1500 and
$N_{\theta}=100$. We found that the location depends very weakly on
$N_{\theta}$ as long as it is larger than 100. On the other hand, the
convergence of the location is slow with increasing $N_r$
(cf. Figure 1). However, with $N_r \agt 1000$, the convergence is good, 
except for the region near mass-shedding limits.

In the typical computations, 
the equatorial radius of a star is covered by $0.6N_r$ grid points. 
In our method, the outer boundaries are located at a finite radius, 
and thus, the outer boundary conditions are approximate. 
However, the stellar equatorial radius is larger than 
several hundred $M$ for stable rotating stars with $n \geq 3$ 
(cf. Table 2), implying that 
the radius of the outer boundaries is larger than $\sim 1000M$. 
In this setting, even the approximate condition is very accurate. 
To check the magnitude of the numerical error associated 
with this approximation, we performed several computations
changing the location of the outer boundaries with a fixed 
grid spacing, and indeed found that the errors in mass, density, and
angular momentum are much smaller than $1\%$.

Figure 1 is a summary of the numerical results for $\Gamma=1.328$,
1.329, 1.330, 1.332, 1.333, and 4/3.
In each panel, the dashed curve denotes 
the sequence of rotating stars at mass-shedding limits.
Namely, there is no rigidly rotating star
in the right-hand side of these curves. 
The solid curve denotes the sequences of constant values of $J$.
The thick solid curve, the crosses, the solid squares, and the open 
circles with the dotted curve are sequences
of the marginally stable stars with 
$N_r=1000$, 500, 750, and 1500, respectively.
It is found that there is a critical value of $J=J_{\rm min}$, 
where $J_{\rm min}$ denotes a constant,  
below which all the stars are unstable.
The values of $J_{\rm min}$ are determined at the intersection
between the curves for the sequence of the mass-shedding limit
and for the sequence of the turning points (cf. Table 2).
In the following, we refer to such a point as an intersection point. 

Here, we write the density of the marginally stable
stars as $\rho_t(J)$ for $J \geq J_{\rm min}$. 
From the fact that all the nonrotating stars are unstable 
for $\Gamma \leq 4/3$, 
we can determine that with $\rho < \rho_t$ for 
a given value of $J \geq J_{\rm min}$, the stars are unstable and otherwise,
they are stable. Therefore, the rotating stars located 
between the sequences of the mass-shedding limits and
of the turning points are stable against 
quasi-radial oscillations for $\Gamma < 4/3$. 
For $\Gamma=4/3$, stars with 
$\rho > \rho_t$ for a given value of $J$ are unstable. 
Thus, the stars located in the higher-mass side of 
the sequence of the turning points are stable. 

The long-dashed curves denote the relation of equation (\ref{eq24})
with $K=1$; the sequence of the marginally stable
stars derived by the semi-analytic calculation in \S 2.
It is found that the long-dashed curves agree approximately
with the numerical sequences of the turning points
for $\Gamma \geq 1.329$. The error in 
mass for a given value of the central density is typically $0.1\%$. 
This indicates that the semi-analytic calculation gives a good
approximate solution for the marginally stable stars. 

For $\Gamma=1.328$, the long-dashed curve is located
in the right-hand side (the higher-density side) 
of the dashed curve, where there are no rigidly rotating stars.
Since in the semi-analytic calculation the configuration of the stars
cannot be determined, unrealistic solutions for marginally stable
stars may be derived. 
Thus, the long-dashed curve denotes the unrealistic sequence of
the marginally stable stars and indicates 
that there are no stable stars for $\Gamma=1.328$. 

The intersection point 
denotes the most compact marginally stable star for a given equation of state.
Unfortunately, it is not easy 
to identify the turning points accurately near the
mass-shedding limits. This fact can be also recognized
from the fact that convergence of the numerical solutions
for the marginally stable stars with increasing $N_r$
is achieved more slowly near the mass-shedding sequence
than for other solutions. 

Figure 1 shows that 
the numerical sequences of the turning points appear 
to bend sharply near the mass-shedding limits irrespective of
the value of $\Gamma$. However, 
such tendency seems to be spurious because 
with increasing $N_r$, the point of the bending 
tends to approach the mass-shedding sequences and
eventually disappear for $N_r \rightarrow \infty$.
This illustrates that the intersection point is not
accurately determined from the raw data sets for $N_r \leq 1500$. 
In the present work, for the determination
of the intersection point, we pick up the 
data sets of the turning points slightly 
far away from the mass-shedding sequences, and perform
an extrapolation of the turning points to the mass-shedding limits. 
The several quantities for the intersection point 
determined by this method are listed in Table 2. 
We estimate that the errors of the values for $M$, $J$, $\rho_c$, and $\beta$
are $\sim 0.1\%$, 1\%, 10\%, and 0.1\%, respectively. 

Note that our results for $n=3$ slightly disagree with those 
by Baumgarte \& Shapiro (1999). According to our results, 
the values of $\rho_c$ and $q$ at the intersection point are 
$\approx 1.1 \times 10^{-8}$ and $0.91$, while their results
are $\approx 0.7 \times 10^{-8}$ and $0.97$, respectively.
This disagreement comes partly from 
numerical errors contained in our results of 
magnitude $\sim 10\%$ for $\rho_c$ and $\sim 1\%$ for $q$. However, 
even if this error is taken into account,
the results of the two groups disagree. 
Currently, the reason is not very clear.
Here, we point out a fact: In the limit $\rho_c \rightarrow 0$,
the mass of the marginally stable stars from 
our results converges to $\approx 4.555$, 
which agrees with the exact value, 
the value of the Newtonian polytrope with $n=3$.
On the other hand, the asymptotic value of their results
slightly disagrees with the exact value in this limit.
Thus, as far as 
the results for $\rho_c \rightarrow 0$ are concerned, our
numerical results appear to be more accurate. 

Figure 1 shows that for the smaller value of $\Gamma$, 
the allowed region for the stable stars becomes narrower.
We find that for $\Gamma \leq 1.328$, there is no region
for the stable stars and that for $\Gamma = 1.329$, the region for
the stable stars is very narrow.
From these results, we conclude that
all the rigidly rotating stars are unstable for $\Gamma \alt 1.329$. 

According to a semi-analytic estimate presented in \S 2,
marginally stable stars satisfy equation (\ref{eq26}).
In this relation, the terms $3\Gamma-4$, $2(3\Gamma-5)\beta$, and
$2(3\Gamma-5)k_4k_2^{-1}M^{2/3}\rho_c^{1/3}$ are
of nearly identical order as $\sim 10^{-2}$ (except for the cases
$\Gamma \approx 4/3$ in which $3\Gamma-4 \approx 0$).
To confirm the validity of this relation,
we calculated the relative error defined as
\beq
Q \equiv {1 \over 2(3\Gamma - 5)\beta}\biggl[3\Gamma - 4 -2 (3\Gamma -5)
\biggl(\beta -{k_4 \over k_2}M^{2/3}\rho_c^{1/3}\biggr)\biggr]. \label{eq299}
\eeq
Figure 2 shows the values of
$Q$ as a function of $J$ for $\Gamma=1.329$, 1.330,
1.332, and 1.333. It is found that $|Q|$ is less than $\sim 2$\%
irrespective of $\Gamma$. This indicates fair validity of
the relation (\ref{eq26}). 

For $\Gamma=1.329$--1.332, the value of $Q$ appears to asymptotically
approach a constant with increasing $J$ (with decreasing $\rho_c$).
This is reasonable, because 
even in the limit $\rho_c \rightarrow 0$, a marginally stable
star is rotating, and hence, the semi-analytic calculation in which
the stars are assumed to be spherical contains certain systematic
error. Therefore, the asymptotic value of $Q$ may be regarded as
a typical magnitude of the systematic error associated with
neglect of the nonspherical deformation in the semi-analytic calculation. 

Equation (\ref{eqqq}) implies that if $q_{\rm min}$ is larger than
unity, the value of $q$ for all the marginally stable stars is
larger than unity. As shown above and in Table 2, $q_{\rm min}$ 
is larger than unity for $1.329 \alt \Gamma \alt 1.332$, implying 
that the value of $q$ is always so for the rigidly rotating and
marginally stable stars. 

\section{Predicting the final outcome}

The marginally stable stars determined in \S 3 may be plausible 
approximate initial conditions for a 
stellar core collapse or a pair-unstable 
collapse. Here, we predict the outcome of the collapse
paying particular attention to the black hole formation case
under the assumptions that 
(i) the collapse proceeds in an axisymmetric manner,
(ii) the viscous angular momentum transport during the
collapse is negligible, and (iii) 
the pressure and heating effect never halt the collapse. 
Actually, in a stellar core collapse with mass larger than
$\sim 40 M_{\odot}$ (Fryer 1999) and in a pair-unstable collapse
with mass larger than $\sim 260 M_{\odot}$ (Fryer et al. 2001),
the pressure support and the nuclear burning energy are 
so small that the final outcome is likely to be a black hole. 
The numerical analysis is carried out in the same manner as that of 
Shibata \& Shapiro (2002). 

Since viscosity is assumed to be negligible during the collapse, 
the specific angular momentum $j$ of each fluid element is conserved
in an axisymmetric system. Here, $j$ is defined as 
\beq
j \equiv h u_{\varphi}. 
\eeq
Next, we define rest-mass distribution $m_*(j)$ as a function of $j$, 
which is the integrated baryon rest-mass of all fluid 
elements with specific angular momentum less than a given value $j_0$:  
\beq
m_*(j_0) \equiv 2\pi \int_{j < j_0} \rho u^t B e^{2\zeta-2\nu} 
r^2 dr d(\cos\theta). 
\eeq
Let us assume that a seed black hole is formed during the collapse and
consider the innermost stable circular orbit (ISCO) around the growing
black hole at the center. If $j$ of a fluid element is smaller than
the value at the ISCO $(j_{\rm ISCO})$,
the element will fall into the black hole eventually.
Now the possibility exists that some fluid can be captured 
even for $j > j_{\rm ISCO}$, if it is in a noncircular orbit. 
Ignoring these trajectories  yields the minimum amount of mass 
that will fall into the black hole at each moment. 
The value of $j_{\rm ISCO}$ changes as the black hole grows. 
If $j_{\rm ISCO}$ increases, additional mass  
will fall into the black hole. However,
if $j_{\rm ISCO}$ decreases, ambient fluid 
will no longer be captured. This expectation 
suggests that when $j_{\rm ISCO}$ reaches a maximum value, 
the dynamical growth of the black hole will have already terminated
(i.e., before the maximum, the point for $j=j_{\rm ISCO}$ will be
reached). 
As Shibata \& Shapiro (2002) demonstrated, prediction of 
the growth of the mass and spin by this method is in good agreement
with a numerical result in the collapse with $n=3$, 
suggesting that this is also a good method for $n \agt 3$. 

To analyze the growth of the black hole mass, we generate Figures 3 and 4. 
Figure 3 shows $m_*(j)/M_*$ as a function of $j/M_*$ 
for $\Gamma=1.329$, 1.330, 1.332, and 1.333. 
Here, we choose the marginally stable rotating stars
at mass-shedding limits. That is, 
we choose the most compact and most rapidly rotating, 
marginally stable stars. 

Each panel (i) in Figures 4 (a)--(d) denotes 
$q_*\equiv J(j)/m_*(j)^2$ as a function of $m_*(j)/M_*$. Here, 
$J(j)$ denotes the total angular momentum 
with the specific angular momentum less than a 
given value $j_0$ and is defined according to 
\beqn
J(j_0)=2\pi \int_{j < j_0} \rho h u^t u_{\varphi} B e^{2\zeta-2\nu}
r^2 dr d(\cos\theta). 
\eeqn
Now, $J(j)/m_*(j)^2$ and $m_*(j)$ may be approximately 
regarded as the instantaneous spin parameter and mass of a black hole,
since the baryon rest-mass is nearly equal to the gravitational mass
for the soft equations of state.
Therefore, the solid curve in each panel (i) of Figure 4 
may be interpreted as an approximate evolutionary track 
for the angular momentum parameter of the growing black hole. 
It indicates that with increasing the black hole mass, 
the spin parameter also increases. 

If we assume that $m_*(j)$ and $q_*$ are the  
instantaneous mass and spin parameter of 
the growing black hole and that the spacetime 
can be approximated instantaneously by a Kerr metric, 
we can compute $j_{\rm ISCO}$  
(Bardeen et al. 1972; chapter 12 of Shapiro \& Teukolsky, 1983). 
In each panel (ii) of Figure 4, we show $j_{\rm ISCO}[m_*(j),q_*(j)]$ 
as a function of $m_*(j)$.
The maximum of $j_{\rm ISCO}$ is reached at 
$m_*(j) = m_{*\rm crit}$, where $m_{*\rm crit}/M_* \approx 0.57$,
0.77, 0.89, and 0.92 for $\Gamma=1.329$, 1.330, 1.332, and 1.333, 
respectively (Fig. 4, {\it circles}). The value of 
$j_{\rm ISCO}$ is slightly smaller than $j$ at $m_*(j) = m_{*\rm crit}$: 
The point for $j=j_{\rm ISCO}$ 
(Fig. 4, {\it triangles}) is reached at a value of $m_*(j)$, which
is only slightly smaller than $m_{*\rm crit}$. 
This suggests that the mass of the black hole will increase to
$\sim m_{*\rm crit}$. However, 
after the maximum of $j_{\rm ISCO}$ is reached, 
$j_{\rm ISCO}/m_*(j)$ steeply decreases above this mass fraction. 
Thus, once the black hole reaches the point of $m_*(j) = m_{*\rm crit}$,
it will stop growing dynamically. 
Panels (ii) of Figure 4 shows that 
at this stage, $q_* \approx 0.94$, 0.87, 0.77, and 0.73 for 
$\Gamma=1.329$, 1.330, 1.332, and 1.333, respectively. 

It is interesting to note that for $\Gamma=1.329$--1.332, 
the total value of $q$ for the system is larger than unity. However,
as far as the inner region of the collapsing stars is concerned,
it is much smaller than unity and, hence, a seed black hole is likely
to be formed. The large value of $q>1$ will be reflected 
in the large mass fraction of formed
disks surrounding the central black hole. 
For the cases with $\Gamma=1.329$, 1.330, and 1.332, the fraction of
the disk mass will be $\sim 40$\%, 20\%, and 10\%, respectively. 
The present numerical analysis indicates that the global
value of $q$ is not always a good indicator for predicting
the final outcome of stellar collapse with soft equations of state.
However, the conclusion drawn here is based on
several assumptions. To confirm it more strictly, 
fully general relativistic simulations are necessary.

After the dynamical collapse, a system of a black hole and
surrounding disks in a nearly quasiequilibrium state will be formed.
Subsequent evolution of the system is determined by the 
viscous time scale of the accretion disks. Since 
a large fraction of the total mass forms 
disks that eventually fall into the central black hole, 
the black hole will spin up to $q \alt 1$ for $\Gamma \alt 1.332$. 

Next, we pay attention to the fate of the collapse for
general marginally-stable stars. 
For the analysis, we only need to repeat the same procedure described 
above. Fortunately, without detailed numerical computations, 
it is possible to approximately
predict the outcome using a semi-analytic calculation as follows.

In the semi-analytic method adopted in \S 2,
the mass and angular momentum contained inside a cylinder of
radius $\varpi$ are written as
\beqn
M(\varpi)=M {\cal M}(y),\\
J(\varpi)=J {\cal J}(y),
\eeqn
where $y \equiv \varpi \xi_1/R_s$
and $\xi_1$ denotes the Lane-Emden radial coordinate of the stellar surface
(e.g., Shapiro \& Teukolsky 1983).  
${\cal M}$ and ${\cal J}$ are computed from the Lane-Emden
functions as
\beqn
&&{\cal M}(y)={\xi_1 \over \theta'_1}
\int_0^y \theta^n \xi^2 (1-\cos t)d\xi,\\
&&{\cal J}(y)={\xi_1 \over \kappa\theta'_1}\int_0^y \theta^n \xi^4
\Big(1-\cos t-{1-\cos^3t \over 3}\Big)d\xi,~~~~~~~~
\eeqn
where $t=\sin^{-1}({\rm min}[1, \varpi \xi_1/ (R_s \xi)])$, $\theta'_1$
denotes $|d\theta/d\xi|$ at $\xi=\xi_1$, and
\beq
\kappa \equiv {2 \over 3\xi_1^4\theta'_1}\int_0^{\xi_1} \theta^n\xi^4d\xi.
\eeq

Now we define the nondimensional angular momentum parameter
estimated for the fluid elements inside the cylinder of
radius $\varpi$ as
\beqn
q(\varpi) \equiv  {J(\varpi) \over M(\varpi)^2} 
= q  {{\cal J} \over {\cal M}^2}. \label{eqeq}
\eeqn
Thus, a curve of $q(\varpi)$ is determined for a given value of $q$
in this model. 

To verify that this semi-analytic estimate for
$q(\varpi)$ is a good approximation, we plot $q(\varpi)$ as
a function of ${\cal M}$ in each panel (i) of Figure 4 (dotted curves).
To plot these curves, we choose the same value of $q$
for each corresponding numerical model presented in Figure 4.
Note that the specific angular momentum is written as $\varpi^2 \Omega$
in this semi-analytic model. Thus, $M(\varpi)$ may be regarded as
the mass distribution as a function of the specific angular momentum.

Comparison between the dotted and solid curves in Figure 4 
confirms that the numerical and semi-analytic results are in 
good agreement. 
This indicates that the semi-analytic method adopted here is 
appropriate for an approximately quantitative study. 

According to equation (\ref{eqeq}), the curves of the semi-analytic results
shift upward with increasing $q$. For a sufficiently large 
value of $q$, the value of $q(\varpi=0)$ exceeds unity and, 
therefore, $q(\varpi) > 1$ for any value of $\varpi$. 
Such large values of $q$ are achieved for marginally stable stars 
of small compactness. In this case, it is natural to expect that
even a seed black hole will not be formed in the gravitational collapse. 
Thus, the collapse of marginally stable stars with
$1.329 \leq \Gamma < 4/3$ and with a small compactness will not result
directly in a black hole but in a disk or a torus. 
The critical value of $q$ above which a black hole will not be formed 
($q_{\rm BH}$) is $\sim 2.5$ irrespective of $\Gamma$
(see the solid and dotted curves for $q=2.517$ in Figure 4(a)
for an example). 

Using equations (\ref{eq24}), (\ref{eqqq}), and (\ref{eq27}), the
compactness of the marginally stable stars can be written as a
function of $q$: 
\beq
{M \over R_s}={k_2(3-n)\alpha \over 2k_5(3-2n)(q^2 -k_4/k_5)}. 
\eeq
For $q=2.5$, $M/R_s=2.61\times 10^{-4}$, $2.02 \times 10^{-4}$,
$8.18 \times 10^{-5}$, and $2.06 \times 10^{-5}$ for
$\Gamma=1.329$, 1.330, 1.332, and 1.333, respectively.
In the iron core collapse of massive stars,
the radius and mass before the collapse are $\sim$ a few thousand kilometer
and at most a few $M_{\odot}$, respectively
(Umeda \& Nomoto, unpublished). 
Thus, the compactness is $\sim 10^{-3}$ and, hence, 
the value of $q$ at the onset of the instability
is likely to be smaller than 2.5 if the star is assumed to be rigidly
rotating. On the other hand, in the pair-unstable collapse,
the density and mass at the onset of the collapse are
$\sim 10^4~{\rm g/cm^3}$ and $100 M_{\odot}$
(Bond et al. 1984). This implies that the compactness 
is $\sim 2\times 10^{-4}$. At the onset of a pair-unstable collapse, the
adiabatic index of the star decreases from $\sim 4/3$ to smaller 
value on a time scale much 
longer than the dynamical time scale (Bond et al. 1984).
If the star is rotating sufficiently rapidly, 
the instability will set in when $\Gamma$ decreases below $\sim 1.329$. 
For $\Gamma=1.329$ with $M/R_s \sim 2\times 10^{-4}$, the value of 
$q$ at the onset of the instability is larger than 2.5
if the star is rigidly rotating.
This suggests that the collapse may not lead to a black hole
directly in this case. 

A collapse with $q > q_{\rm BH}$ will result in formation of a disk or a
torus, which will be subsequently unstable against nonaxisymmetric 
deformation. After the nonaxisymmetric instabilities turn on,
angular momentum will be transported from the inner region to
the outer region, decreasing the value of $q$ around the inner region
below unity and resulting in the formation of a seed black hole. 

To clarify whether the above scenario is correct, it is obviously necessary 
to perform a three-dimensional numerical simulation in general 
relativity (e.g., Shibata 1999; Shibata \& Ury\=u 2000, 2002; 
Font et al. 2002; Shibata et al. 2003). 
During the collapse, the typical length scale may change 
by a factor of $10^4$ from $R_s / M \sim 10^4$ to 1. 
To follow the collapse by numerical simulation, 
very large computational resources will be necessary and, thus, 
the simulation for this phenomenon will be one of 
the computational challenges in the field of numerical relativity. 

Finally, we note the following point. 
As shown above, $q(\varpi)$ near the rotational axis is 
smaller than the global value of $q$ by a factor of 2 
in the rigidly rotating case. This is the reason that the relation 
$q>1$ is not likely to be a good criterion for no black hole formation.
However, in the differentially rotating case or 
for stiffer equations of state with $\Gamma \agt 2$,
the ratio $q/q(\varpi=0)$ becomes smaller than 2. 
For $q/q(\varpi=0)\agt 1$, the global value of $q$ may be still an
approximate indicator
for predicting the outcome. 
Previous axisymmetric simulations in general relativity have not
been performed taking an equilibrium with
$\Gamma \sim 4/3$ in the rigidly rotating 
initial condition (Nakamura 1981; Stark \& Piran 1985; 
Piran \& Stark 1986; Nakamura et al. 1987; Shibata 2000). 
This is the reason why the previous numerical works have suggested
that the value of $q$ is a good indicator for predicting the outcome. 

\section{Summary} 

We have investigated the secular stability of rigidly rotating stars against
a quasi-radial oscillation in general relativity. It is found that
all the rigidly rotating stars with $\Gamma \leq 1.328$ are 
unstable against a quasi-radial oscillation. 
Stars with $1.329 \alt \Gamma < 4/3$ that are unstable 
for the spherical case can be stabilized by the 
effect of rotation. Therefore, for rapidly rotating stars
for which the adiabatic index gradually decreases from $\sim 4/3$
to smaller values, the instability against gravitational collapse will set in
when the value of $\Gamma$ decreases below $\approx 1.329$ (not 4/3). 

The marginally stable stars for $\Gamma=1.329$--4/3 have been determined
numerically. It is found that for $1.329 \leq \Gamma \leq 1.332$, 
the nondimensional angular momentum parameter $q$ for {\it all} 
rigidly rotating and marginally stable stars is larger than unity. 
The value of $q$ for the marginally stable stars
increases with decrease of the compactness. Therefore, 
for a sufficiently small compactness, the value of 
$q$ exceeds unity even for $1.332 < \Gamma < 4/3$. 
These results are in good agreement with those derived 
by a simple semi-analytic calculation presented in \S 2. 

The final outcomes of axisymmetric collapse of the 
rigidly rotating and marginally stable
stars are predicted assuming that the pressure never
halts the collapse and that the mass distribution as a function of the specific
angular momentum is preserved. It is found that 
even for the case $q > 1$, a black hole will form as a
result of the gravitational collapse 
if the value of $q$ is smaller than $\sim 2.5$. This is due to 
the fact that the angular momentum in the 
central region of the rigidly rotating stars is not large 
enough and that the effective value of the nondimensional angular momentum
parameter $q$ for the inner region is smaller than unity.
The outcome of the collapse in such cases 
will be a rapidly rotating black hole surrounded by a massive disk. 
For $q \agt 2.5$, even the central region has angular momentum large
enough to prevent collapsing to a black hole. In this case, 
the outcome will be a disk or a torus. It will be subsequently unstable 
against nonaxisymmetric deformation.
Once the nonaxisymmetric structure is developed,
the angular momentum will be transported outward
by the nonaxisymmetric effects, 
decreasing the value of $q$ in the inner region below unity and
eventually forming a seed black hole. 

Although the conjecture mentioned above is reasonable, 
the final fate of the collapse of marginally 
stable stars with $q \agt 1$ can be determined only by fully 
general relativistic simulations. 
In particular for $q \agt 2.5$, three-dimensional simulations are
necessary since the collapsing star is likely to become
unstable against nonaxisymmetric deformation. 
Since the length scale will change by a factor of $10^4$ during
the collapse, huge computational resources will be necessary 
for the simulation. The simulation for this problem will be one of
the computational challenges in the field of numerical relativity. 

\acknowledgments

Numerical computations were in part performed 
on the FACOM VPP5000 machine in the data processing center of 
National Astronomical Observatory of Japan. 
This work is in part supported by Japanese Monbu-Kagakusho Grants 
14047207, 15037204, and 15740142.


\begin{table}[t]
\begin{center}
\caption{Structure constants of spherical polytropes}
\begin{tabular}{cccccc}
\tableline\tableline
$\Gamma$ & $k_1$ & $k_2$ & $k_4$ & $k_5$ & $\alpha$ \\ \tableline
1.328 & 1.7877 & 0.63546 & 0.92247 & 1.1837 & 0.41330
\\ \tableline 
1.329 & 1.7816 & 0.63613 & 0.92169 & 1.1876 & 0.41571
\\ \tableline
1.330 & 1.7756 & 0.63680 & 0.92091 & 1.1914 & 0.41810 
\\ \tableline
1.332 & 1.7637 & 0.63812 & 0.91934 & 1.1991 & 0.42285
\\ \tableline
1.333 & 1.7578 & 0.63878 & 0.91856 & 1.2028 & 0.42522
\\ \tableline 
4/3   & 1.7558 & 0.63900 & 0.91829 & 1.2041 & 0.42600
\\ \tableline
\end{tabular}
\tablecomments{Values of $k_i$ and $\alpha$ are computed from
the Lane-Emden function. 
}
\end{center}
\end{table}

\begin{table}[t]
\begin{center}
\caption{Parameters of marginally stable stars at mass-shedding limits}
\begin{tabular}{ccccccc}
\tableline\tableline
$\Gamma$ & $M$ & $J$ & $q$ & $\beta$ & $R/M$ & $\rho_c$
\\ \tableline
1.329  & 5.435 & 52.0 & 1.76  & 0.00847 & 2.2e3 & 1.3e-10 
\\ \tableline
1.330  & 5.179 & 37.0 & 1.38  & 0.00856 & 1.4e3 & 6.3e-10 
\\ \tableline
1.332  & 4.779 & 23.6 & 1.03  & 0.00873 & 7.2e2 & 4.5e-9
\\ \tableline
1.333  & 4.613 & 20.0 & 0.940 & 0.00883 & 6.0e2 & 8.8e-9 
\\ \tableline
4/3    & 4.560 & 19.0 & 0.914 & 0.00886 & 5.6e2 & 1.1e-8
\\ \tableline
\end{tabular}
\tablecomments{Adiabatic index, 
$M$, $J$, $q=J/M^2$, $\beta=T/|W|$, $R/M$, and $\rho_c$
of marginally stable stars at mass-shedding limits are shown 
with the units $c=G=K=1$. Note that $M_*$ is nearly equal to $M$
in all the cases (the relative difference is smaller than $10^{-4}$). 
}
\end{center}
\end{table}

\begin{figure}[htb]
\begin{center}
\epsfxsize=2.8in
\leavevmode
(a)\epsffile{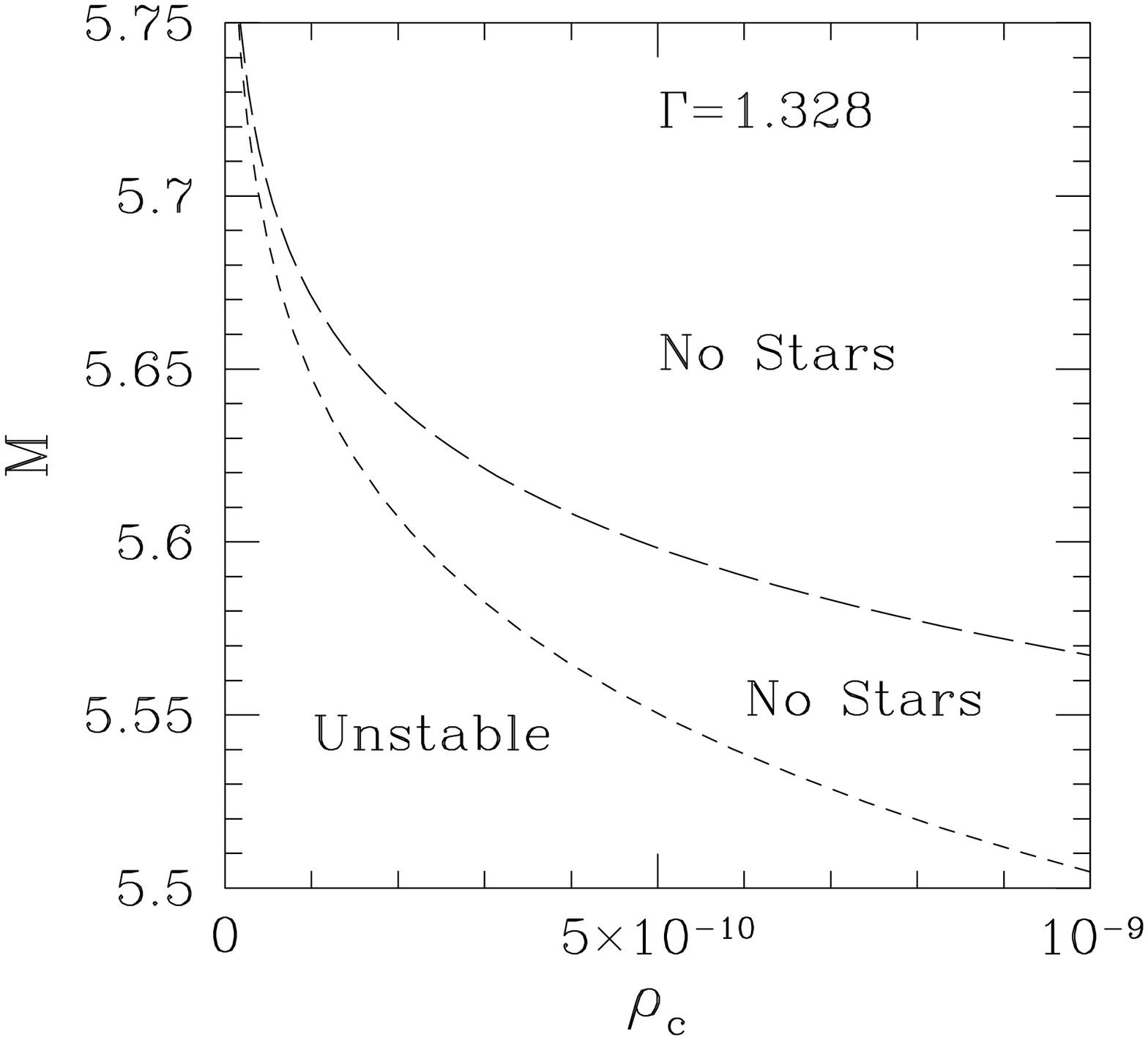}
\epsfxsize=2.8in
\leavevmode
(b)\epsffile{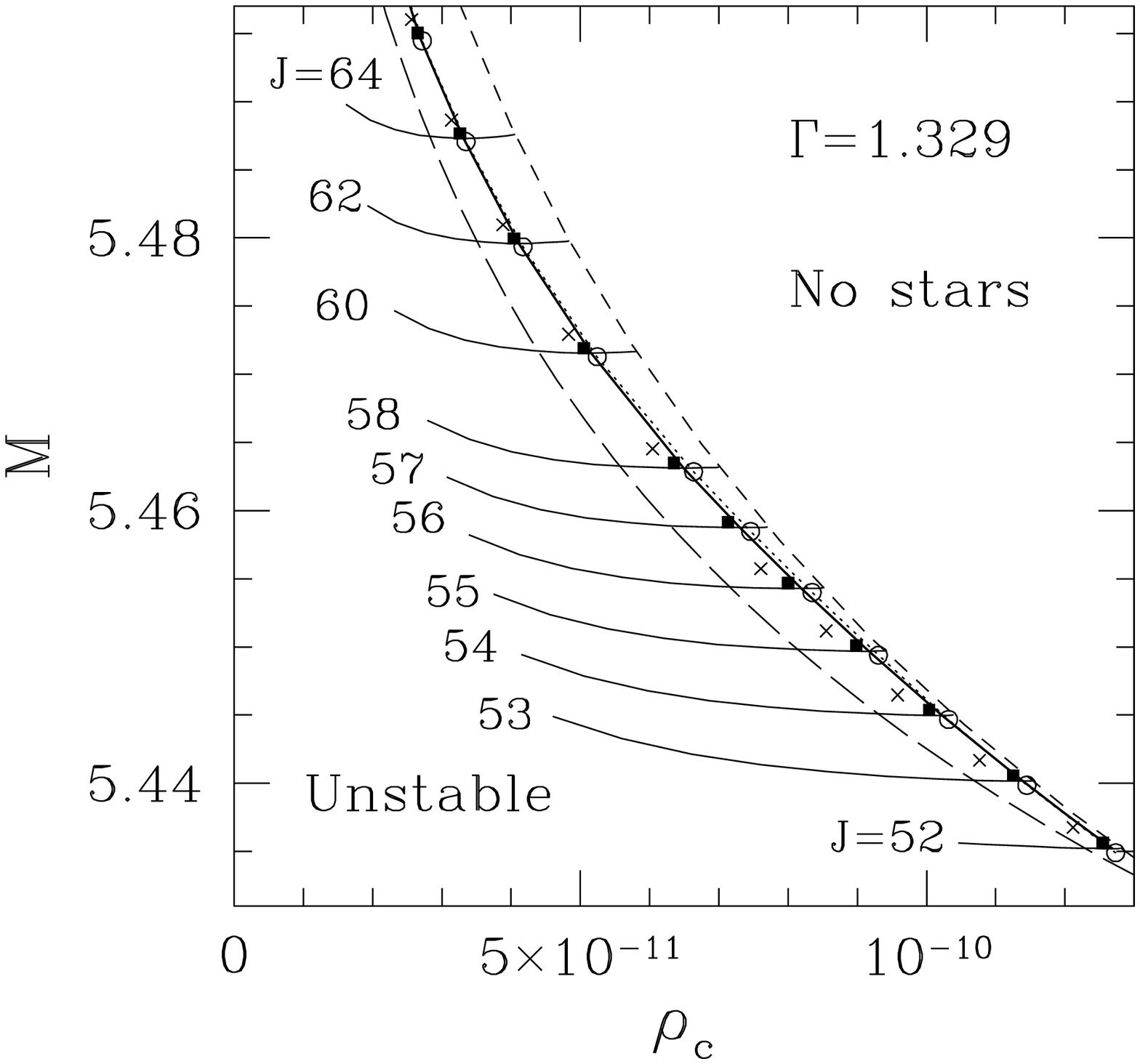}\\
\epsfxsize=2.8in
\leavevmode
(c)\epsffile{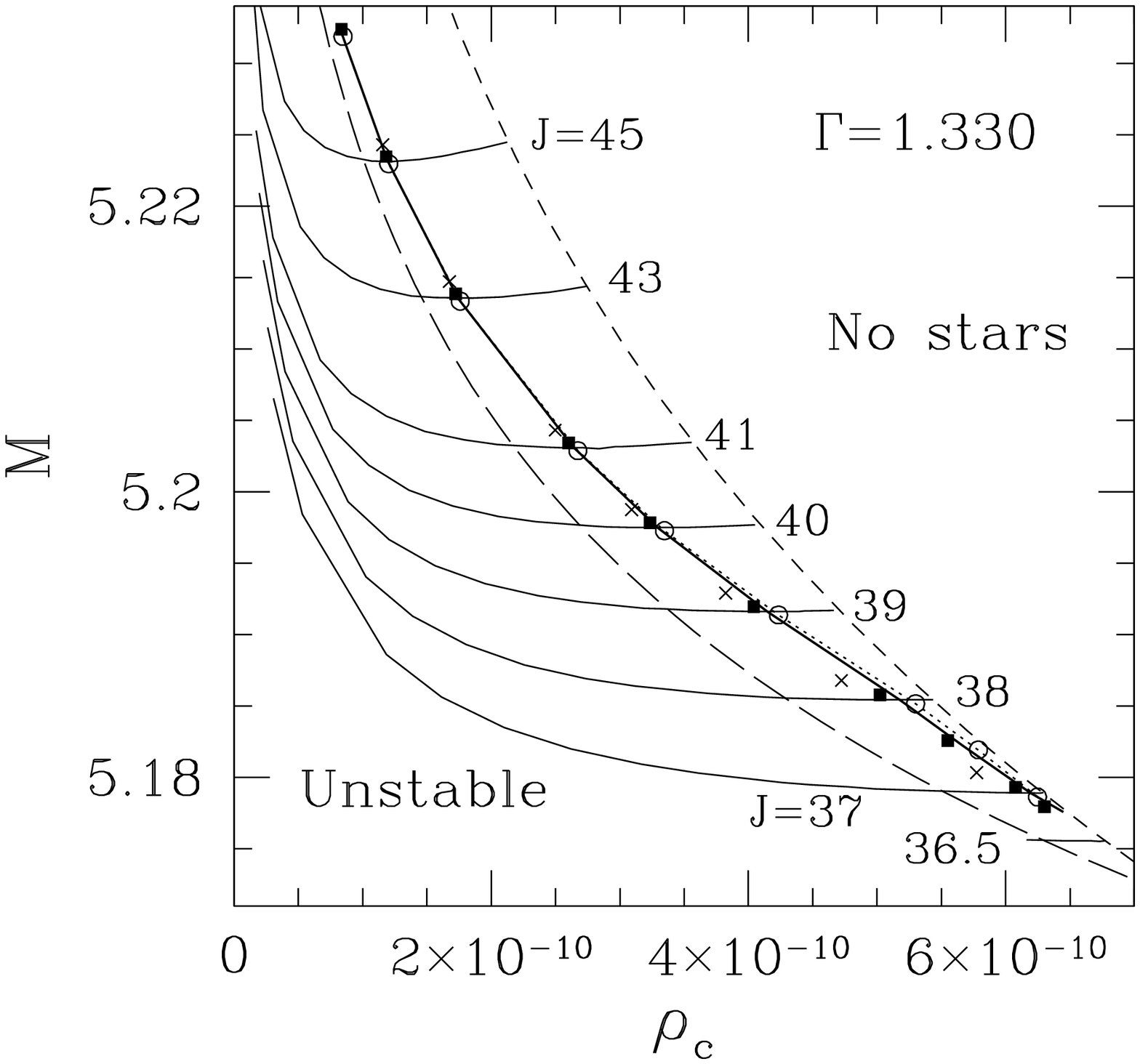}
\epsfxsize=2.8in
\leavevmode
(d)\epsffile{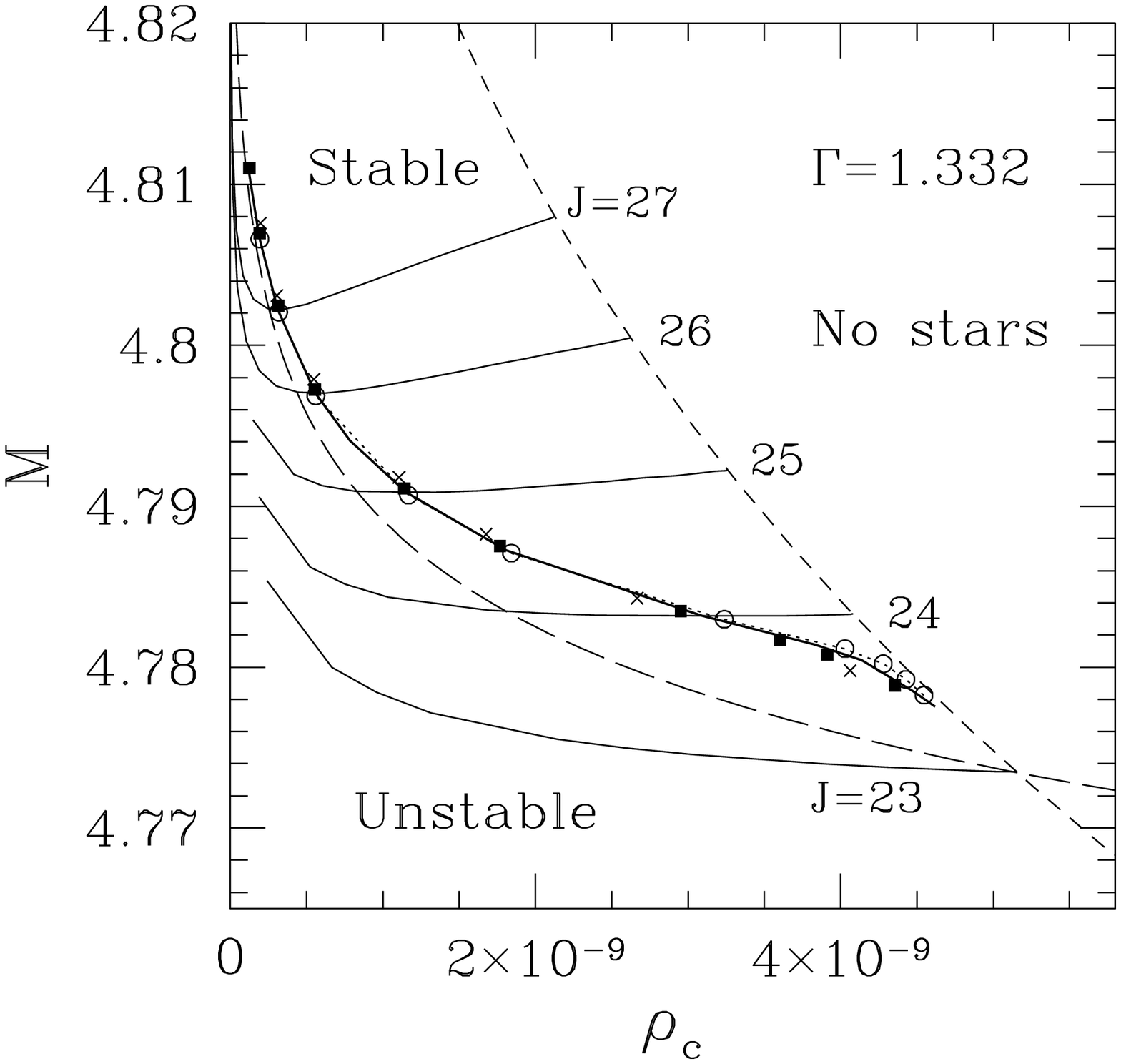}\\
\epsfxsize=2.8in
\leavevmode
(e)\epsffile{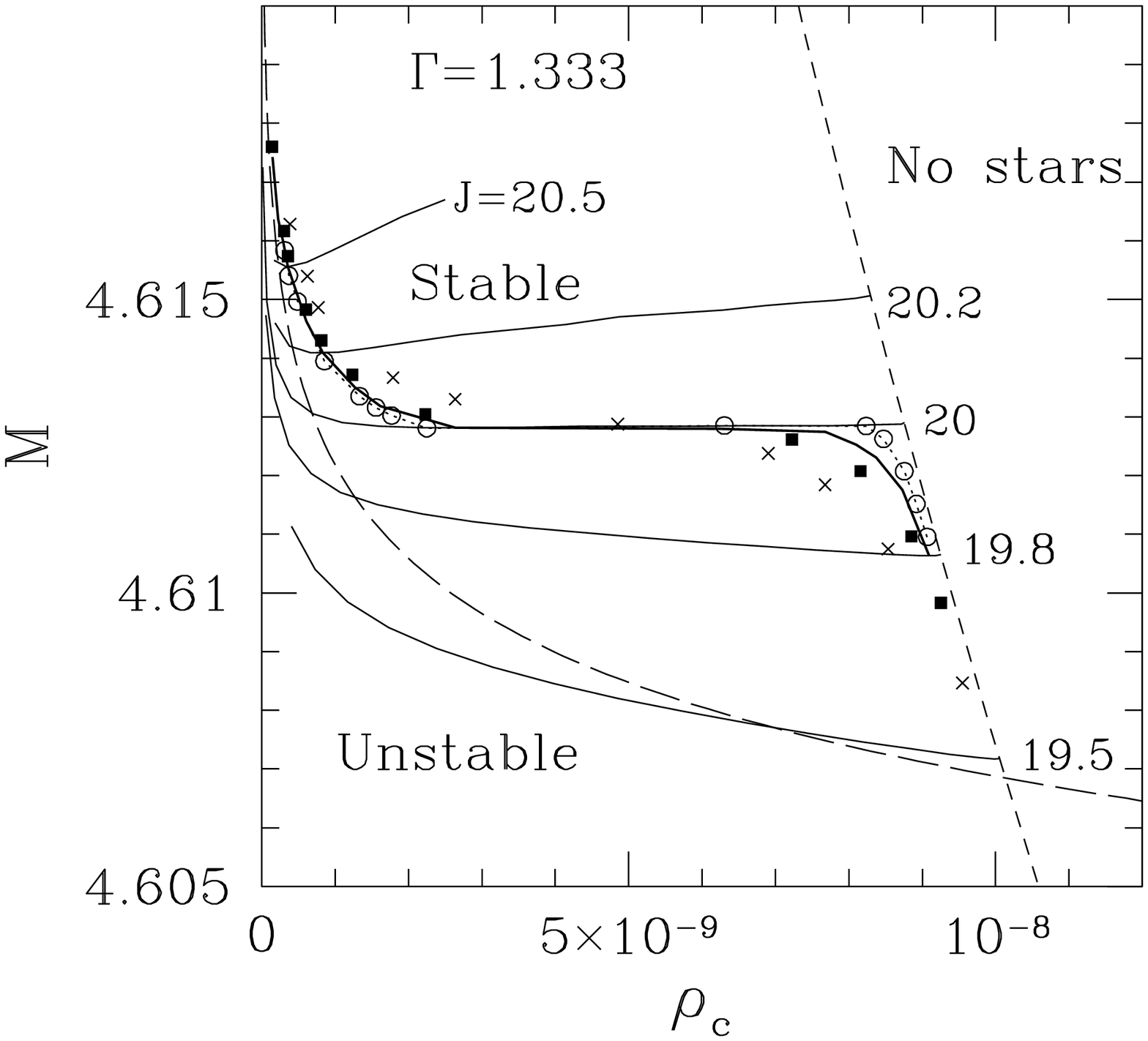}
\epsfxsize=2.8in
\leavevmode
(f)\epsffile{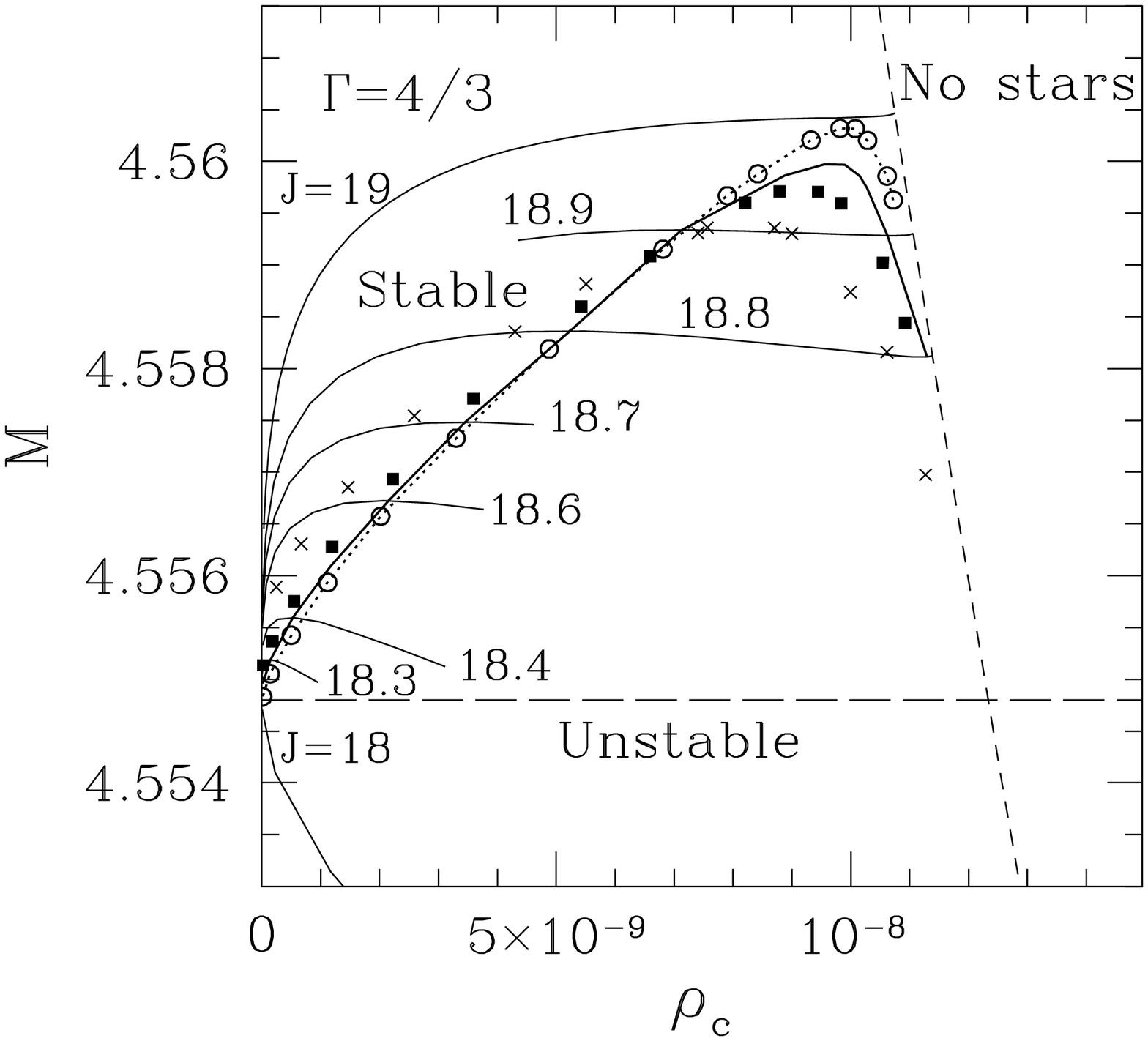}\\
\caption{Gravitational mass 
$M$ as a function of $\rho_c$ for fixed values of $J$
(thin solid curves), for sequences of rotating stars
at mass-shedding limits (dashed curves), 
and for sequences of marginally stable stars
(thick solid curves)
for $\Gamma=1.328$, 1.329, 1.330, 1.332, 1.333, and 4/3 with $N_r=1000$. 
Rigidly rotating stars are located on the left-hand side of
the dashed curves, and thus, 
for $\Gamma=1.328$, all the rigidly rotating stars are unstable.
The long-dashed curves denote the sequences of the marginally stable stars 
derived by a semi-analytic calculation in \S 2.
The crosses, filled squares, and the dotted curves with the open circles
denote sequences of the marginally stable stars
computed with $N_r=500$, 750, and 1500, respectively. 
\label{FIG1}
}
\end{center}
\end{figure}

\begin{figure}[htb]
\begin{center}
\epsfxsize=4.5in
\leavevmode
\epsffile{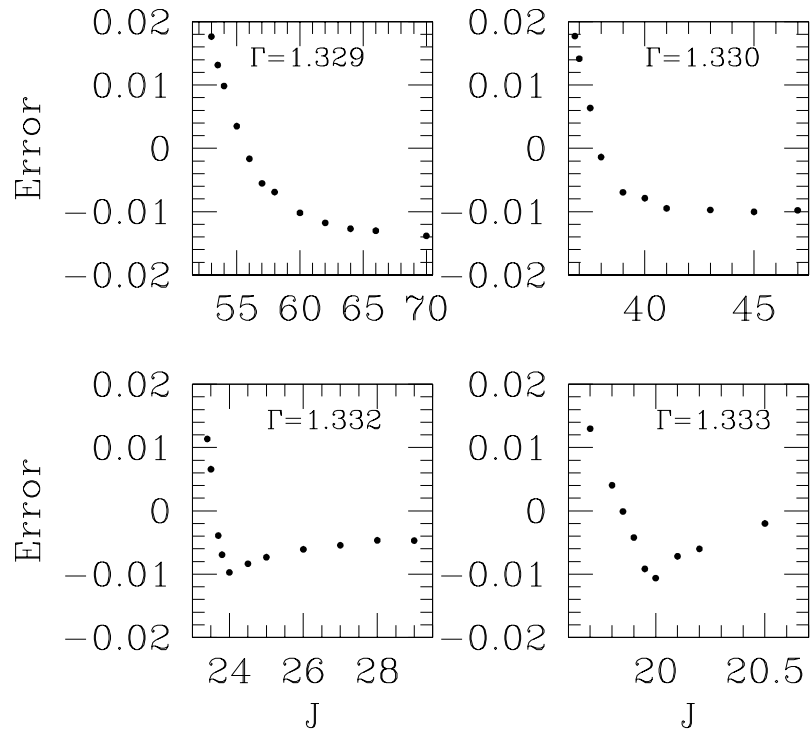}
\caption{Relative error $Q$ for marginally stable stars defined
in equation (\ref{eq299}) as a function of $J$ for 
$\Gamma=1.329$, 1.330, 1.332, and 1.333. 
\label{FIG2}
}
\end{center}
\end{figure}

\begin{figure}[htb]
\begin{center}
\epsfxsize=4.5in
\leavevmode
\epsffile{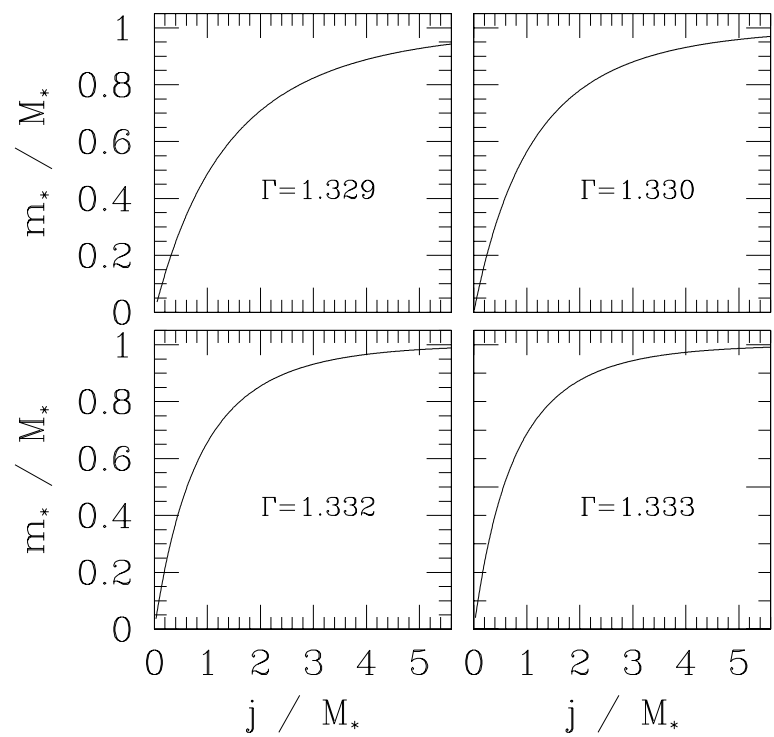}
\caption{Fraction of the rest mass $m_*(j)/M_*$ as a function of $j/M_*$
for the marginally stable stars at mass-shedding limits with 
$\Gamma=1.329$, 1.330, 1.332, and 1.333. 
($J=52.0$, 37.0, 23.6, and 20.0, respectively; see Figure 1.) 
\label{FIG3}
}
\end{center}
\end{figure}

\begin{figure}[htb]
\begin{center}
\epsfxsize=2.9in
\leavevmode
(a)\epsffile{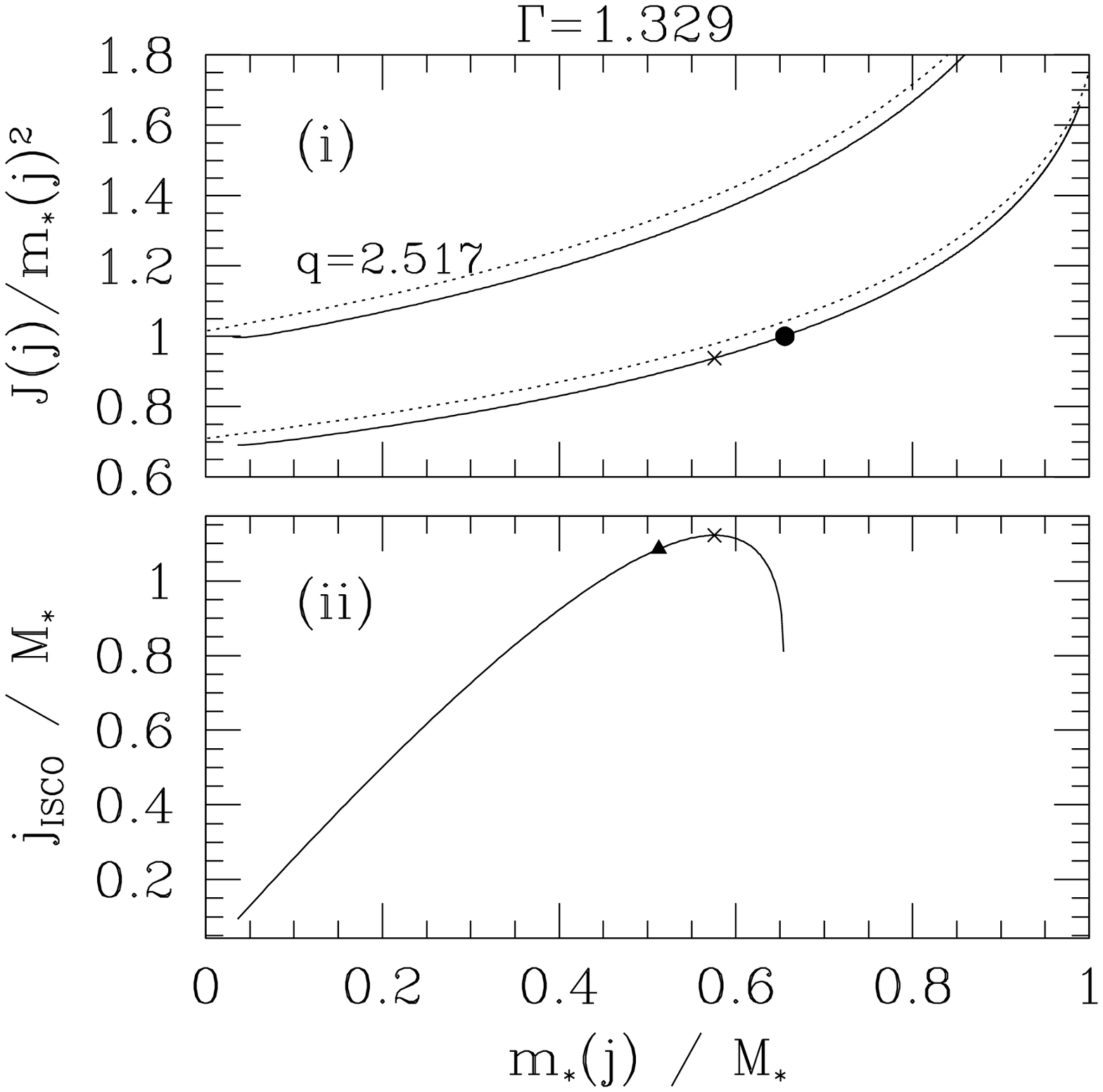}
\epsfxsize=2.9in
\leavevmode
~~~(b)\epsffile{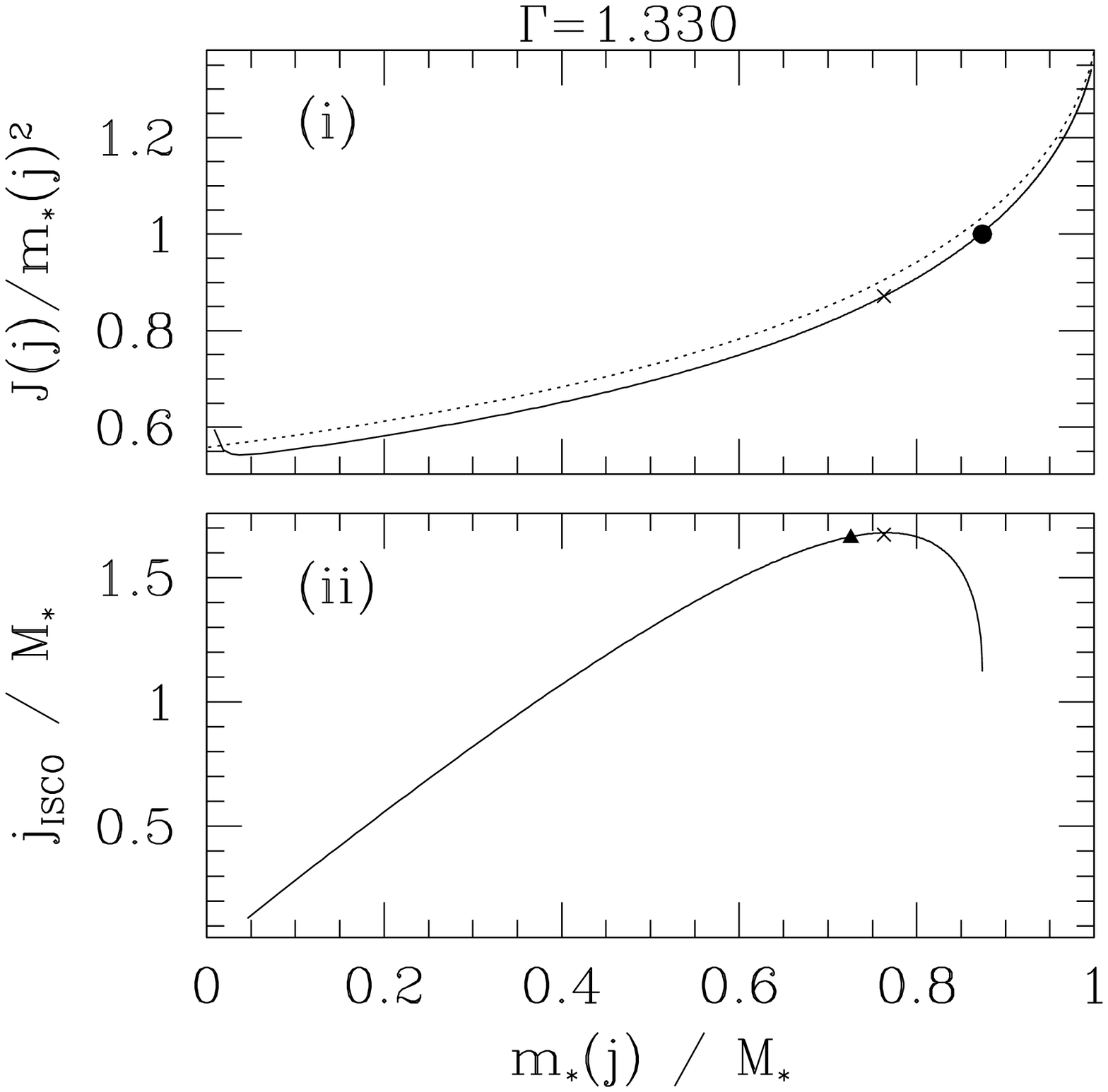}\\
\epsfxsize=2.9in
\leavevmode
(c)\epsffile{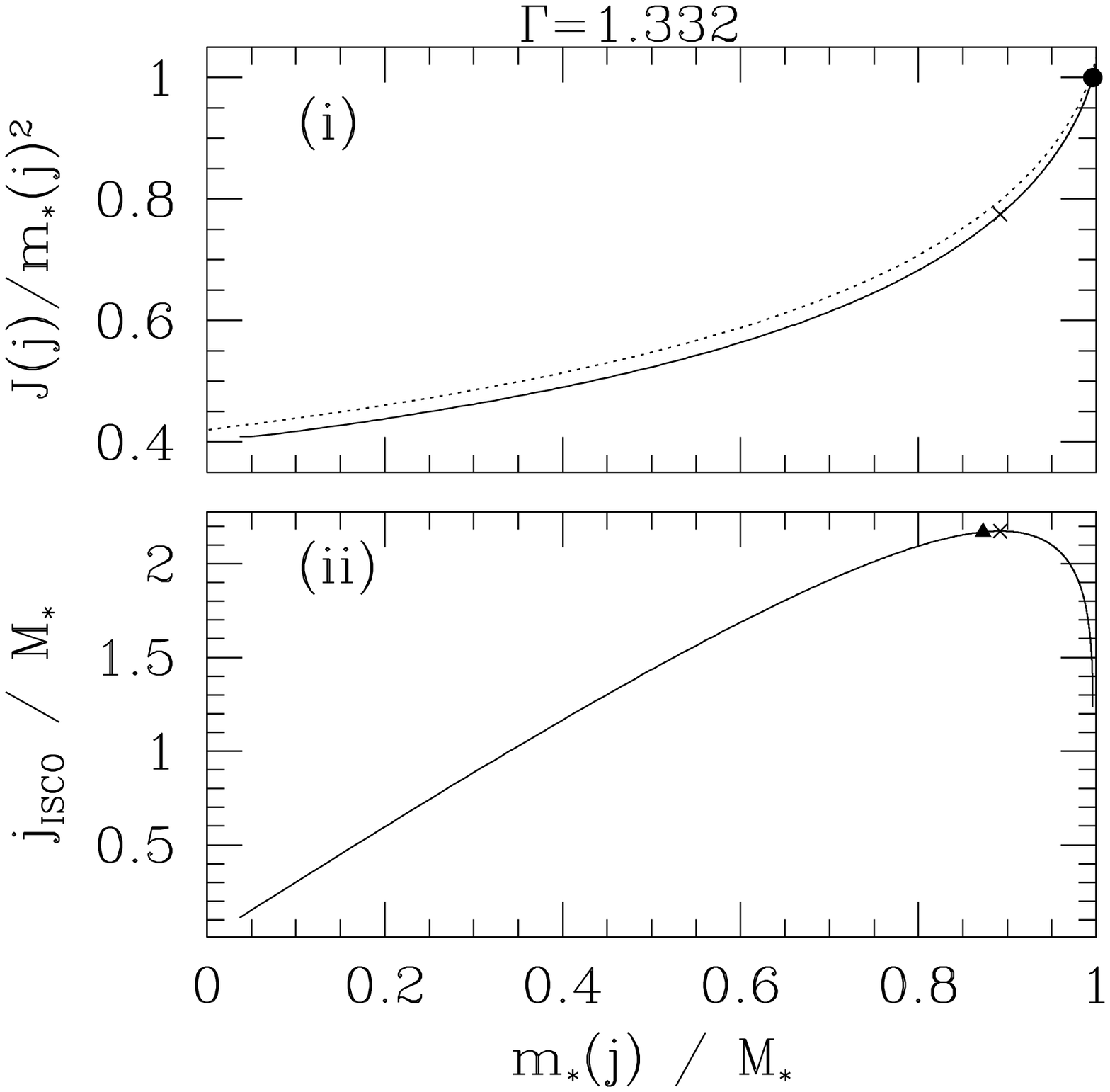}
\epsfxsize=2.9in
\leavevmode
~~~(d)\epsffile{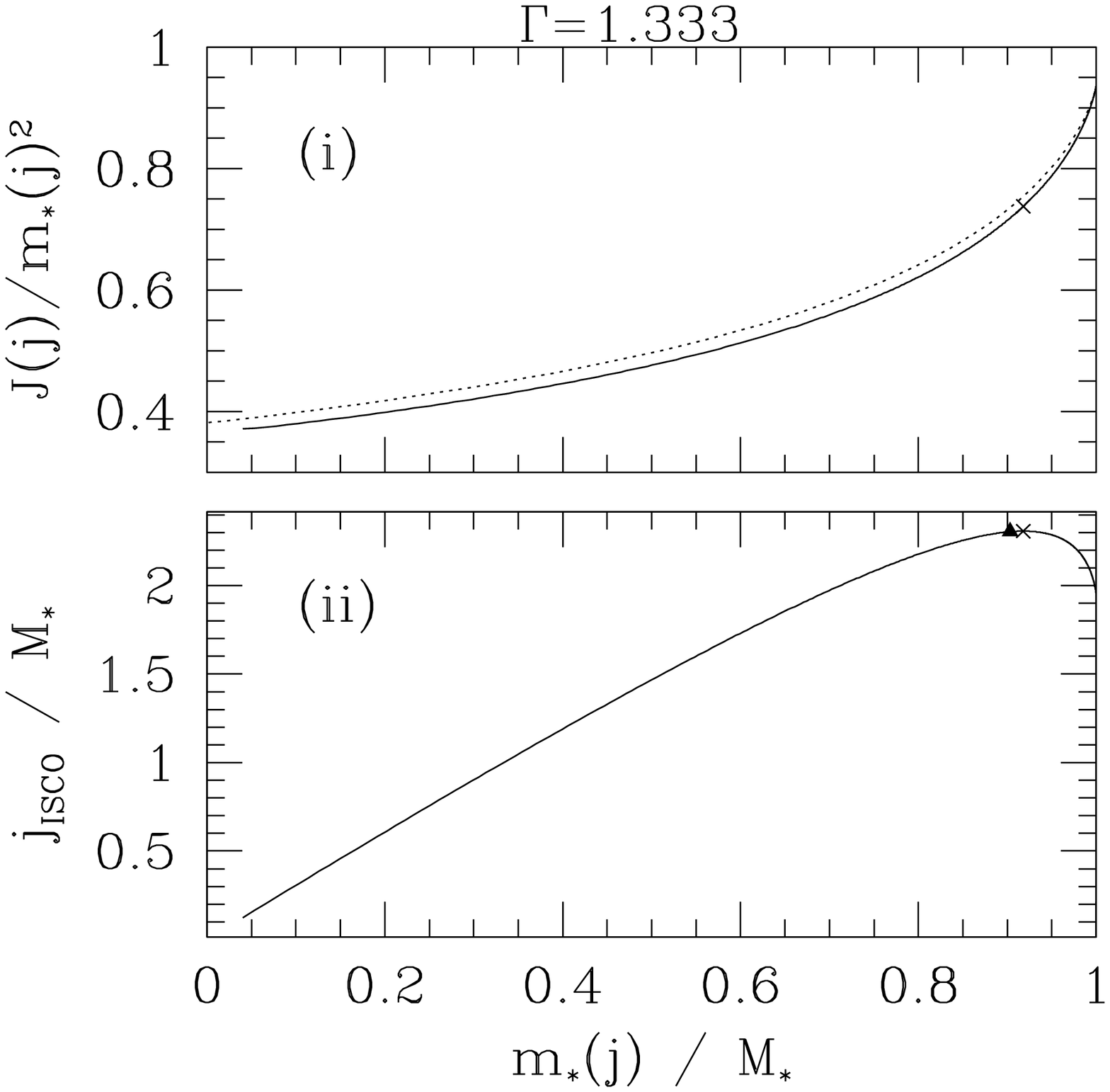}\\
\caption{$q_*\equiv J(j)/m_*(j)^2$ and $j_{\rm ISCO}/M_*$ as a function of
$m_*(j)/M_*$ 
for rigidly rotating stars at mass-shedding limits
near the marginally stable criterion.
(a) $\Gamma=1.329$ and $J=52$ $(q=1.76)$,
(b) $\Gamma=1.330$ and $J=37.0$ ($q=1.38)$,
(c) $\Gamma=1.332$ and $J=23.6$ ($q=1.03$),
and (d) $\Gamma=1.333$ and $J=20.0$ ($q=0.94$).
The crosses denote the point at which $j_{\rm ISCO}$ reaches maximum, 
the filled triangles the point at which $j=j_{\rm ISCO}$, 
and the filled circles the point at which $q_*=1$. 
The dotted curves denote the semi-analytic results for 
the same value of $q$ as in the numerical results. 
For $\Gamma=1.329$, the numerical and semi-analytic results
for $q=2.517$, for which $q(\varpi) \geq 1$ and hence 
a black hole will not be formed, are also plotted. 
\label{FIG4}
}
\end{center}
\end{figure}


\end{document}